\magnification=1200
\baselineskip24truept

The physics of kHz QPOs---strong gravity's coupled anharmonic oscillators

W{\l}odek Klu{\'z}niak$^{1,2,3}$ and Marek Artur Abramowicz$^{1,3,4}$

$^1$Nordita, Blegdamsvej 17, DK-2100 Copenhagen, Denmark

$^2$Institute of Astronomy ``Johannes Kepler'', University of Zielona G\'ora,
Poland

$^3$Institut d'Astrophysique de Paris, 98bis Boulevard Arago, F 75014 Paris,
France

$^4$Department of Astronomy and Astrophysics, Chalmers University,
412-96 G\"oteborg, Sweden

\bigskip
Abstract

We explain the origin of the puzzling high frequency peaks (QPOs) in the
variability power spectra of accreting neutron stars and black holes as
a non-linear 1:2 or 1:3 resonance between orbital and radial epicyclic
motion. These resonances are present because the gravitational field
deviates strongly from a Newtonian $1/r$ potential. Our theory agrees
with the recently reported observations of two QPOs, at 300 Hz and 450
Hz, in the black hole candidate 1655-40.

\vskip2truecm

\vfill\eject

Observations of accreting neutron stars in low-mass X-ray binaries
reveal two preferred frequencies, none of which is fixed. These
frequencies, called kHz QPOs (quasi-periodic oscillations), show up as
peaks in the observed power spectrum (the Fourier transform of the time
variation) in the X-ray flux, [1]. Their properties are a major puzzle.

The peaks typically come in pairs, at frequencies $\omega_1$ and
$\omega_2$ both on the order of a kHz ($\times 2\pi$), but both varying
considerably in any given source (by several hundred Hertz in intervals
of hours), with their difference $\omega_1-\omega_2$ showing markedly
less variation. That the difference is not constant (as would be
expected if one of the two observed frequencies were a beat between the
other frequency and the constant rotation rate of the star) is a clear
indication that two fundamental frequencies are present, none of which
is the stellar spin frequency.

Another puzzle has been that in black hole candidates only one high
frequency QPO had been reported, with properties similar to one of the
variable kHz QPOs in neutron stars. Observations of the black hole
candidate 1655-40, reported after the original submission date of this
Letter$^\ddagger$, show that in fact two high frequencies show up also in
black-hole candidates.

Newtonian gravity with the $1/r^2$ force law is scale-free, there is no
preferred frequency (Fig. 1). If gravity were so described, only two
fundamental frequencies would be expected for a quasi-spherical star with
a thin accretion disk [2]: the stellar rotational frequency $\Omega_*$,
and the Keplerian orbital frequency at the surface of the star
$\Omega_K(R)=(GM/R^3)^{1/2}$. The most important fact about these two
frequencies is that they are fixed for a star of fixed mass $M$, radius
$R$, and angular momentum.

It has long been recognized [3] that for black holes general relativity
predicts instead two other preferred frequencies, also fixed for a given
gravitating body: the orbital frequency in the innermost (marginally)
stable orbit, $\Omega(r_{ms})$, and the maximum epicyclic frequency
$\omega_{max}= {\rm max}(\omega_r)\equiv \omega_r(r_{max}) $, (Fig. 2).
These reflect the presence of a characteristic scale, the gravitational
radius $r_g= 2GM/c^2$. For example, in the Schwarzschild metric
$r_{ms}=3r_g$, $r_{max}=4r_g$, $\Omega=\Omega_K$, and
$\omega_{max}=\Omega(4r_g)/2$. (The importance of the maximum in the
epicyclic frequency was first stressed in a seminal paper by Kato and
Fukue [4].) Thus, strong gravity presents us with two frequencies, but
these are fixed for a given star and hence cannot be identified with the
observed variable kHz QPOs in neutron stars.
 
However, we note that in contrast with Newtonian gravity of spherical
bodies where the only frequency at a given radius is $\Omega_K(r)$, in
general relativity turbulent noise may excite epicyclic motions at the
different frequency $\omega_r$, so inhomogeneities in flow at radius $r$
contribute to the power spectrum at (angular) frequencies $\Omega$, and
$\omega_r$, as well as at combination frequencies characteristic of
coupled anharmonic oscillators (including rational fractions of
eigenfrequencies), giving a rich structure to the power spectrum of
X-ray variability, in agreement with observations.

We point out that in general relativity, in addition to the fixed
frequencies $\Omega_*$, $\Omega(r_{ms})$, $\omega_r(r_{max})$, there are
other preferred frequencies, those of 1:2, 1:3, etc., resonances between
orbital and radial epicyclic frequencies. These are possible because the
ratio of orbital and radial epicyclic frequencies tends to large values
near the marginally stable orbit:
$\Omega(r)/\omega_r(r)\rightarrow\infty$, as $r\rightarrow r_{ms}$ (Fig.
2).  Frequencies in 1:2 or 1:3 ratio can be in resonance because epicyclic
motion is anharmonic. As is usual for non-linear oscillators, the
resonance occurs for a range of frequencies near the eigenfrequency of
the oscillator, so the driving frequency (the orbital frequency here)
need not be an exact multiple of the eigenfrequency of the epicyclic
oscillator, nor need it be constant [5]. Thus, the resonant frequencies
have just the properties which seemed puzzling in the power spectra of
accreting neutron stars. We suggest that the kHz QPOs are caused by such
resonances, and hence are manifestations of strong-field gravity.

Specifically, we suggest that one of the observed high QPO frequencies
could be one of the variable orbital frequencies driving the 1:2, or 1:3
epicyclic resonance. All other things being equal, the most prominent
peaks are expected where most of the luminosity is generated, and that
is between the marginally stable orbit, at $r=r_{ms}$, where the disk
terminates and the X-ray flux vanishes, and the radius $r_{max}$ where
$\omega_r$ peaks. For the Schwarzschild metric the position (radius) of
the 1:2 resonance, $r_2$, coincides with the epicyclic maximum,
$r_2=r_{max}$, and for rotating bodies $r_2>r_{max}$, typically. For
realistic metrics of rapidly rotating neutron stars, the radius of the
1:3 resonance, $r_3$, is close to $r_{max}$. We expect the prominent
resonance in the power spectrum to be the one closest to $r_{max}$ on
the side of the star, so for rapidly rotating neutron stars the most
prominent resonance should be 1:3, i.e., the one with eigenfrequency
$\omega_r(r_3)=\Omega(r_3)/3$.

For maximally rotating Kerr metric, $r_3$ and $r_{max}$ nearly coincide,
but in view of the relatively low accretion rate (per mass) of most
Galactic black-hole candidates, such as 1655-40, it would not be
surprising if their metrics were not maximally rotating Kerr. For
accretion in the Schwarzschild metric not much luminosity is released at
$r_3=(9/8)r_{ms}$. Thus, there should be much less power in the black
hole QPO, than is the case for neutron stars, as indeed is observed.

We might remark that observations reveal similar electromagnetic spectra
of the X-ray emissions of black hole candidates and of neutron stars, at
least in some states. This strongly suggests that accretion disks in
neutron stars are similar to those in black holes, as they would be if
$R<r_{ms}$, as preferred by our model. The fact that only one high
frequency QPO is observed in black hole candidates has been puzzling.
After this Letter$^\ddagger$ was submitted, the discovery of a second high
frequency peak in the black-hole candidate 1655-40 has been reported
[6]. If the two frequencies are the orbital frequency and its
beat with the epicyclic frequency in 1:3 resonance, as suggested by us,
they should be (approximately) in 3:2 ratio. The reported frequencies in
1655-40 are about 450 Hz and 300 Hz.

In summary, strong-field effects of general relativity, and in
particular metric properties of space-time around rapidly rotating
neutron stars, make natural the excitation of a 1:3 or 1:2 anharmonic
epicyclic resonance, driven by orbital motion whose variable (orbital)
frequency may be imprinted on the X-ray flux as a fairly prominent kHz
QPO. A second QPO may be present at a frequency differing from the first
by the epicyclic frequency of the same resonance, with the difference
frequency varying to a lesser degree than the QPO frequency. The same
high frequency QPOs may appear in black hole systems, and indeed they
now have been reported in the black hole candidate 1655-40 ([6]).

The authors appreciate the hospitality and generous support of Nordita,
where most of this work has been done.

\vfill\eject

REFERENCES AND NOTES

{$^\ddagger$ This manuscript was submitted to Physical Review Letters in
December 2000, before the observational discovery of a second
QPO in a black hole candidate has been reported. We
have added on May 3, 2001 a few sentences explaining that the discovery
confirms our theory.}

[1] van der Klis, M., Ann. Rev. Astron. Astrophys. 38, 717 (2000).

[2] It is thought that fluid in accretion disks of small vertical extent
(``thin disks'') approximately follows circular geodesic trajectories.
Observations of water masers in some thin disks in active galactic
nuclei (AGNs) do indeed reveal a Keplerian distribution of velocity for
the water masers, $v_K=(GM/r)^{1/2}$: Miyoshi, M., et al., Nature 373,
127 (1995).

[3] Kato, S., Mineshige, S., and Fukue, J. {\it Black Hole Accretion
Disks} (Kyoto University Press, Kyoto 1998).

[4] Kato, S., Fukue, J. 1980, PASJ, 32, 377. 

[5] Landau, L.D., Lifschitz, E.M. {\it Classical Mechanics}, (Pergamon,
1973).

[6] Strohmayer, T.E., astro-ph/0104487

\vfill\eject

FIGURE CAPTIONS

Figure 1.  The $1/r$ potential of a Newtonian point mass is scale-free.
The orbital frequency in circular orbit and the frequency of epicyclic
motions coincide (a small perturbation of a test particle originally in
circular motion gives rise to periodic motions with the same period as in
the original orbit).

Figure 2.  In general relativity, the frequency of radial epicyclic
motions, $\omega_r$, goes to zero in the marginally stable orbit.  The
epicyclic frequency, $\omega_r$, and the orbital frequency in circular
orbits, $\Omega$, both in units of $c/r_g$, are shown as a function of
the circumferential radius in units of the gravitational radius
$r_g=2GM/c^2$ (after Kato et al., 1998 [3]).  The numerical values are
for the Schwarzschild metric, where the marginally stable orbit is at
$r=3r_g$, but the qualitative features of the frequencies are general.

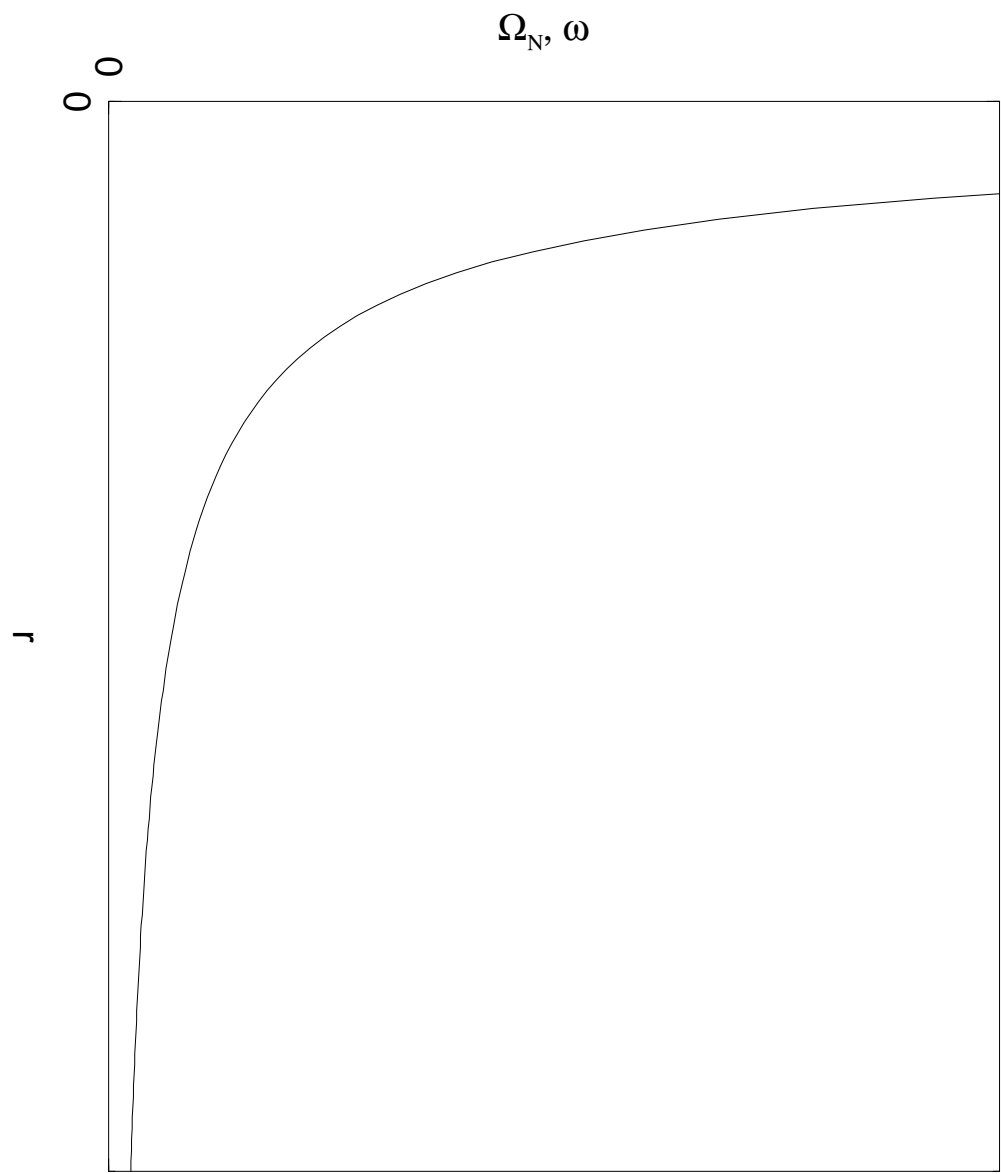
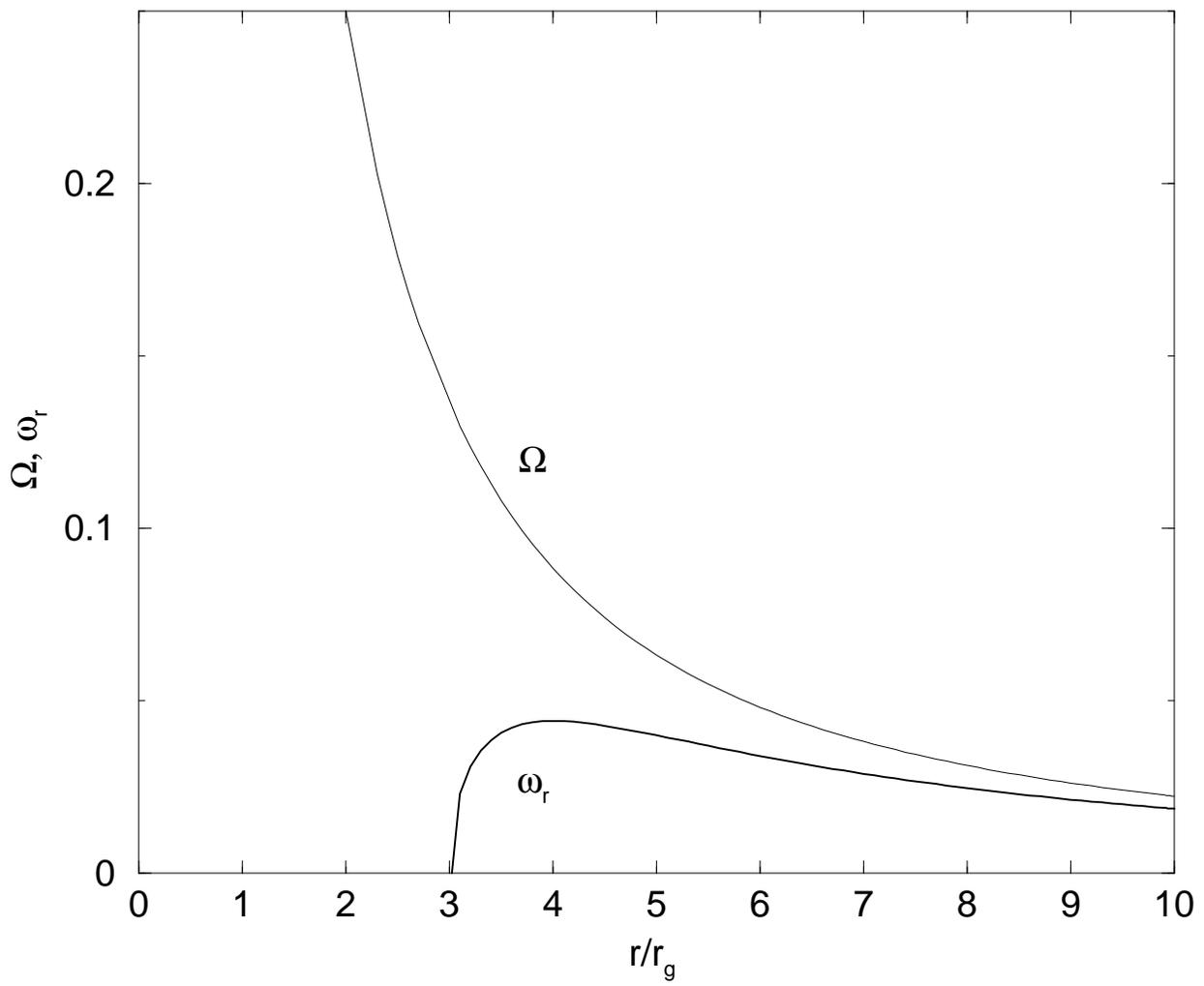
\bye